\documentclass[conference]{IEEEtran}
\IEEEoverridecommandlockouts
\usepackage{cite}
\usepackage{amsmath,amssymb,amsfonts}
\usepackage{graphicx}
\usepackage{textcomp}
\usepackage{xcolor}
\usepackage{multirow, multicol}
\usepackage{mathtools,xparse}
\usepackage[dvips]{epsfig}
\usepackage{adjustbox}
\usepackage{array}
\usepackage{makecell}
\usepackage{mathabx}
\usepackage{caption}
\usepackage{subcaption}
\usepackage{algorithm}
\usepackage{algpseudocode}
\usepackage{tabularx}
\usepackage{float}

\PassOptionsToPackage{hyphens}{url}\usepackage{hyperref}

\def\BibTeX{{\rm B\kern-.05em{\sc i\kern-.025em b}\kern-.08em
    T\kern-.1667em\lower.7ex\hbox{E}\kern-.125emX}}
\begin{document}

\title{Speaker-Adaptive Neural Vocoders for Parametric Speech Synthesis Systems
}

\author{\IEEEauthorblockN{Eunwoo Song$^{1}$, Jin-Seob Kim$^{1}$, Kyungguen Byun$^{2}$ and Hong-Goo Kang$^2$}
\IEEEauthorblockA{\textit{$^1$NAVER Corp., Seongnam, Korea} \\
{\textit{$^2$Yonsei University, Seoul, Korea}}
}
}

\maketitle
\begin{abstract}
	    This paper proposes speaker-adaptive neural vocoders for parametric text-to-speech (TTS) systems. 
	    Recently proposed WaveNet-based neural vocoding systems successfully generate a time sequence of speech signal with an autoregressive framework.
	    However, it remains a challenge to synthesize high-quality speech when the amount of a target speaker's training data is insufficient.
	    %However, it remains a challenge to build high-quality speech synthesis systems when the amount of a target speaker's training data is insufficient.
        To generate more natural speech signals with the constraint of limited training data, we propose a speaker adaptation task with an effective variation of neural vocoding models.
	    In the proposed method, a speaker-independent training method is applied to capture universal attributes embedded in multiple speakers, and the trained model is then optimized to represent the specific characteristics of the target speaker.
	    Experimental results verify that the proposed TTS systems with speaker-adaptive neural vocoders outperform those with traditional source-filter model-based vocoders and those with WaveNet vocoders, trained either speaker-dependently or speaker-independently.
	    In particular, our TTS system achieves 3.80 and 3.77 MOS for the Korean male and Korean female speakers, respectively, even though we use only ten minutes' speech corpus for training the model.
\end{abstract}


	\section{Introduction}\label{sec:intro}

    Waveform generation systems using WaveNet have attracted a great deal of attention in the speech signal processing community thanks to their high quality and ease of use in various applications \cite{van2016conditional, van2016wavenet}.
    In a system of this kind, the time domain speech signal is represented as a sequence of discrete symbols, and its distribution is autoregressively modeled by stacked convolutional neural networks (CNNs). 
    By appropriately conditioning the acoustic features to the input, WaveNet-based systems have also been successfully adopted in a neural vocoder structure for parametric text-to-speech (TTS) systems \cite{tamamori2017speaker, hayashi2017investigation, hu2017ustc, shen2017natural, adiga2018use}.
    
    To further improve the perceptual quality of the synthesized speech, more recent neural excitation vocoders (e.g. ExcitNet \cite{song2019excitnet}) take advantages of the merits from both the parametric LPC vocoder and the WaveNet structure \cite{yoshimura2018mel, juvela2018speaker, hwang2018lp, valin2018lpcnet, tachibana2018investigation}.
    In this framework, an adaptive predictor is used to decouple the formant-related spectral structure from the input speech signal, and the probability distribution of its residual signal (i.e. the excitation signal) is then modeled by the WaveNet network. 
    As variation in the excitation signal is only constrained by vocal cord movement, the training and generation processes become more efficient. 
    As such, TTS systems with the neural excitation vocoders reconstruct more accurate speech signals than the conventional parametric or WaveNet vocoders \cite{song2019excitnet}.

    However, this approach still requires large amounts of training data to faithfully represent the complex mechanics of human speech production. 
    As a result, unnatural outputs are generated when the training data for the target speaker is insufficient (e.g. a database comprising less than ten minutes' speech). 
    The speaker-independent training method that utilizes multiple speakers for a single unified network shows the feasibility of generating diverse characteristics of voices by conditioning the target speaker's acoustic features \cite{hayashi2017investigation}. 
    However, our preliminary experiments verify that this approach still generates discontinuous speech segments if the target speaker's data is not included in the training process. 
    This problem is more prominent under a TTS framework where prediction errors in estimating auxiliary parameters are inevitable; prediction errors are propagated throughout the autoregressive generation process.
    
    To alleviate this problem, we propose a speaker-adaptive training method for neural vocoding systems. 
    In this framework, to address the lack of speaker-specific information caused by limited training data for a target speaker, a model is trained independently of the target speaker such that it extracts universal attributes from multiple speakers \cite{hayashi2017investigation}. 
    This model is then used to initialize the training model of the target speaker, and all weights are fine-tuned to represent the distinctive characteristics within the target's database. 
    Because this adaptation process helps the CNNs capture speaker-specific characteristics, it is also advantageous in reducing the discontinuity problems that occur in conventional speaker-independent models.
    
    We investigate the effectiveness of the proposed method by conducting objective and subjective evaluations with systems designed both dependently and independently of the target speaker. 
    The merits of the proposed method can be found in its robust performance in a pitch modification task because its initial model shares the diverse characteristics extracted from multiple speech databases. 
    Experiments in arbitrary changes to F0 contours confirm that the proposed speaker-adaptive training method synthesizes the modified F0 sound very reliably compared to the conventional speaker-dependent approaches.
    Furthermore, the experimental results show that the proposed method significantly improves the perceptual quality of synthesized speech compared to conventional approaches.

\section{Neural vocoders}
\label{sec:ch3}

%%%%%%%%%%%%%%%%%%% Fig: ExcitNet %%%%%%%%%%%%%%%%%%%%%%%%%%%%%%
	\begin{figure}[!t]
	\begin{minipage}[t]{.495\linewidth}
	\centerline{\epsfig{figure=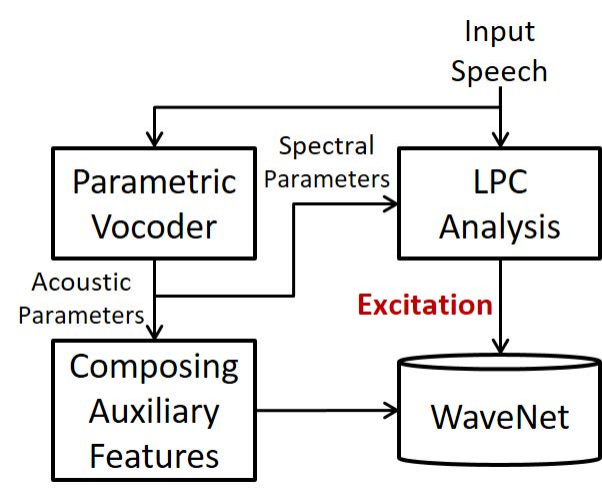,width=40.5mm}}
	%\vspace*{-1pt}	
	\centerline{(a)}  \medskip
	\end{minipage}
	\begin{minipage}[t]{.495\linewidth}
	\centerline{\epsfig{figure=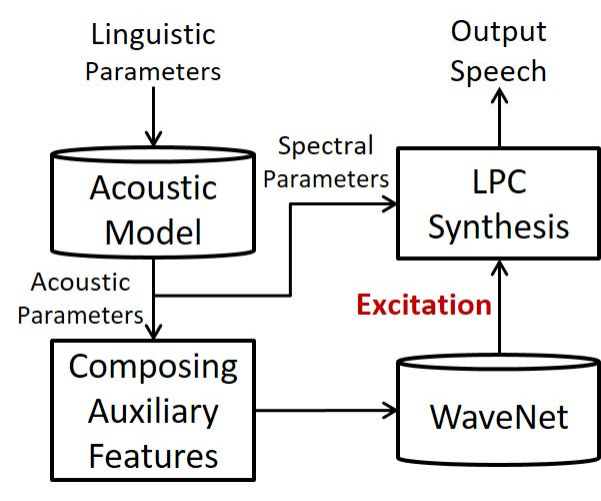,width=40.5mm}}
	%\vspace*{-1pt}	
	\centerline{(b)}  \medskip
	\end{minipage}	
%	\vspace*{-6pt} 
	\caption{ExcitNet vocoder framework for a TTS system: (a) training and (b) synthesis.}	
%	\vspace*{-12pt} 
	\label{fig:ExcitNet}
	\end{figure}
%%%%%%%%%%%%%%%%%%% Fig: ExcitNet %%%%%%%%%%%%%%%%%%%%%%%%%%%%%%

\subsection{WaveNet-based neural vocoding frameworks}
\label{ssec:ch3-1}

    The basic WaveNet framework is an autoregressive network which generates a probability distribution of waveforms from a fixed number of past samples \cite{van2016wavenet}. 
    Recent \textit{WaveNet vocoders} directly utilize acoustic features as the conditional input where these features are extracted from conventional parametric vocoders \cite{tamamori2017speaker, hayashi2017investigation, hu2017ustc, shen2017natural, adiga2018use}.
    This enables the WaveNet system to automatically learn the relationship between acoustic features and speech samples which results in superior perceptual quality over traditional parametric vocoders \cite{tamamori2017speaker, wang2018comparison}.

    However, due to the inherent structural limitations of CNNs in terms of capturing the dynamic nature of speech signals, the WaveNet often generates noisy outputs caused by distortion in the spectral valley regions.
    To improve the quality of synthesized speech, several frequency-dependent noise-shaping filters have been proposed \cite{song2019excitnet, tachibana2018investigation, yoshimura2018mel, juvela2018speaker, hwang2018lp, valin2018lpcnet}.
    In particular, the neural excitation vocoder \textit{ExcitNet} (described in Figure~\ref{fig:ExcitNet}a) exploits a linear prediction (LP)-based adaptive predictor to decouple the spectral formant structure from the input speech signal. 
    The WaveNet-based generation model is then used to train the residual LP component (i.e. the excitation signal).
    %As variation in the excitation signal is only constrained by vocal cord movement, the training process becomes more effective. 

    In the speech synthesis step shown in Figure~\ref{fig:ExcitNet}b, the acoustic parameters of the given input are first generated by an acoustic model designed with a conventional deep learning-based TTS system \cite{song2017effective}.
    Those parameters are used as auxiliary conditional features for the WaveNet model to generate the corresponding time sequence of the excitation signal. 
    Ultimately, the speech signal is reconstructed by passing the generated excitation signal through the LP synthesis filter. 
    %In this way, the quality of the synthesized speech signal is further improved because the spectral component is well represented by the deep learning framework and the residual component is efficiently generated by the WaveNet framework.

\subsection{Speaker-adaptive neural vocoders}
\label{ssec:ch3-3}
    The superiority of neural vocoding systems over traditional parametric vocoders has been explained above but it is still challenging to build a high-quality speech synthesis system when the training data for a target speaker is insufficient, for example with just ten minutes of speech. 
    
\begin{figure}[!t]
    \centerline{\epsfig{figure=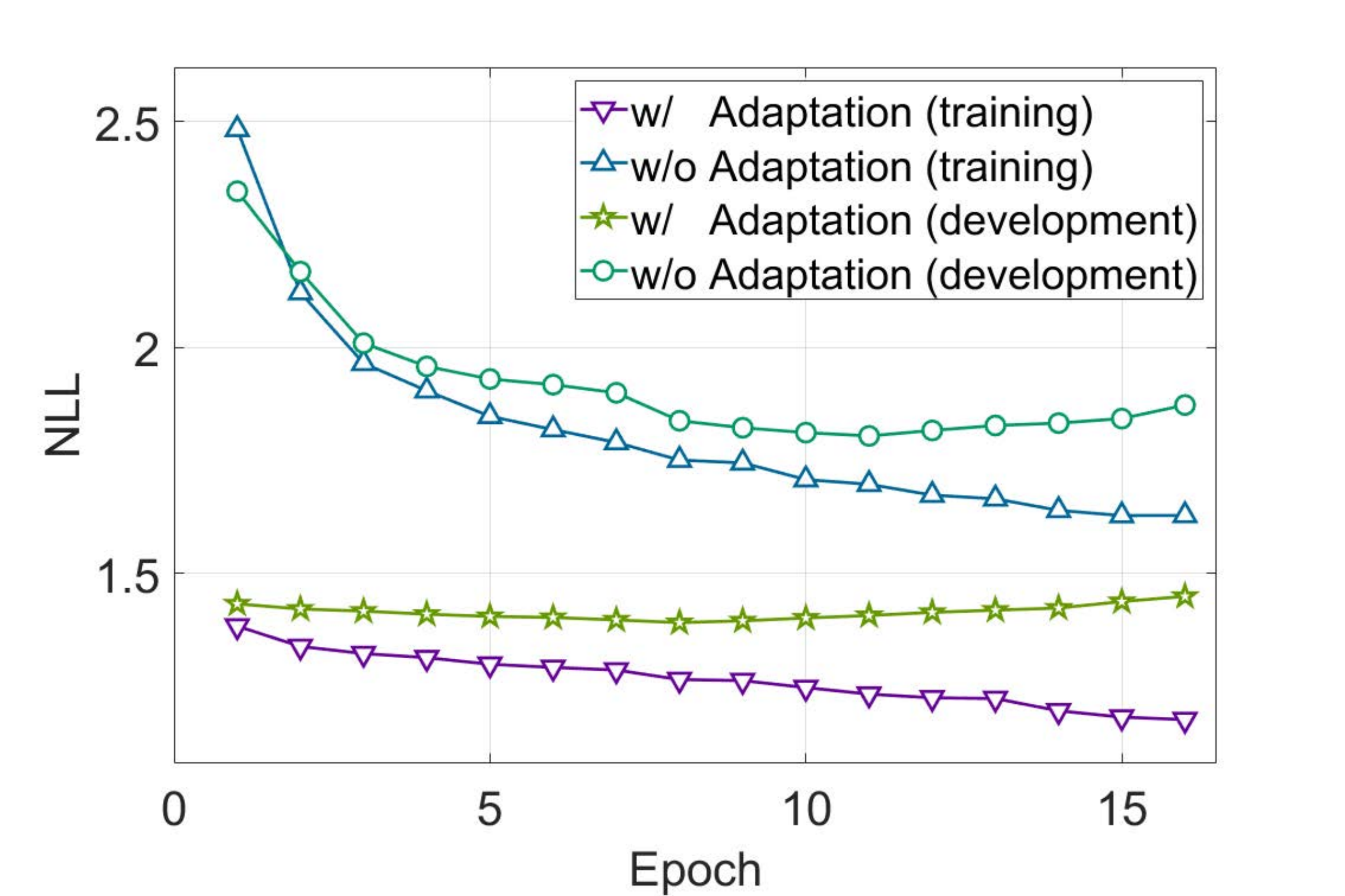,width=67mm}}
%    \vspace*{-2pt}  
    \caption{Negative log-likelihood (NLL) obtained during the training process with (w/) and without (w/o) adaptation.}
    %\vspace*{-16pt}  
    \label{fig:nll}
    \end{figure}

    To generate a more natural speech signal with limited training data, we employ an adaptation task in training the neural vocoders\footnote{
    Note that we only focus on the WaveNet vocoders in this study, but the proposed method can be extended to any of neural vocoders such as RNN- or Glow-based models \cite{mehri2017samplernn, kalchbrenner2018efficient, prenger2019waveglow}.
    }.
    In the proposed framework, a speaker-independently trained multi-speaker model is used as an initializer, and then all weights are updated in training the target speaker's model.
    As the initial model already represents global characteristics embedded in the multiple speakers quite well \cite{hayashi2017investigation}, the fine-tuning mechanism only needs to capture speaker-specific characteristics from the target's data set.
    Consequently, the entire learning process becomes more effective.
    Fig.~\ref{fig:nll} shows the negative log-likelihood obtained during the training phase, of which results confirm that the proposed method significantly reduces both training and development errors as compared to the system without having an adaptation process.

%%%%%%%%%%%%%%%%%%% Fig: NLL %%%%%%%%%%%%%%%%%%%%%%%%%%%%%%  

\section{Experiments}
\label{sec:ch4}

    %%%%%%%%%%%%%%% Table: Dataset %%%%%%%%%%%%%%%%%%%%%%%%
	\begin{table}[!t]   
	\begin{center}         
	\caption{Number of utterances in different sets for the Korean male (KRM) and the Korean female (KRF) speakers (SPKs).}  
%	\vspace*{-4pt}
	\label{table:numUtt}
	{\small        
	\begin{tabular}{>{\centering}m{.20\linewidth}||c|c|c}
	\Xhline{2\arrayrulewidth}
	SPK			& Training  & Development 	& Test \\
			\hline \hline
	KRM 	& 55 (10 min)		& 25 (5 min) 			& 80 (15 min)	\\
			\hline
	KRF		& 90 (10 min)		& 40 (5 min)			& 130 (15 min)	\\
			\Xhline{2\arrayrulewidth}
	\end{tabular}}          
	\end{center}         
%	\vspace*{-16pt}
	\end{table}
    %%%%%%%%%%%%%%% Table: Dataset %%%%%%%%%%%%%%%%%%%%%%%%

\subsection{Experimental setup}
\label{ssec:ch4-1}
    To investigate the effectiveness of the proposed method, we trained neural vocoding models using three different methods:
    \begin{itemize}
    \item {\bf SD}: speaker-dependent training model
    \item {\bf SI}: speaker-independent training model
    \item {\bf SA}: speaker-adaptive training model
    \end{itemize}
    In the SD and SA models, speech corpora recorded by Korean male and Korean female speakers were used. 
    The speech signals were sampled at 24 kHz, and each sample was quantized by 16 bits.
    Table~\ref{table:numUtt} shows the number of utterances in each set.
    To train the SI model, speech corpora recorded by five Korean male and five Korean female speakers not included in training the SD and SA models were used. 
    For this, 6,422 (10 h) and 1,080 (1.7 h) utterances were used for training and development, respectively. 
    The testing set in the SD and SA models was also used to evaluate the SI model.

    To compose the acoustic feature vectors needed for auxiliary input information, the spectral and excitation parameters were extracted using a previously proposed parametric ITFTE vocoder \cite{song2017effective}.
    In this way, 40-dimensional line spectral frequencies (LSFs), 32-dimensional slowly evolving waveform (SEW), 4-dimensional rapidly evolving waveform (REW), the F0, gain, and v\texttt{/}uv were extracted. 
    The frame and shift lengths were set to 20 ms and 5 ms, respectively.

    In the WaveNet training step, all acoustic feature vectors were duplicated from a frame to the samples to match the length of the input speech signals \cite{tamamori2017speaker}. 
    Before training, they were normalized to have zero mean and unit variance. 
    The corresponding speech signal was normalized in in a range between -1.0 and 1.0 and encoded by 8-bit-$\mu$ compression. 
    The WaveNet architecture comprised of three convolutional blocks, each with ten dilated convolution layers with dilations of 1, 2, 4, and so on up to 512. 
    The number of channels of dilated causal convolution and the 1$\times$1 convolution in the residual block were both set to 512. 
    The number of 1$\times$1 convolution channels between the skip-connection and the softmax layer was set to 256. 
    The learning rate was set to 0.0001, and the batch size was set to 30,000 (1.25 sec).
    
    To train the SI-WaveNet model, all data from the multiple number of different speakers were used; the sequence of each batch was randomized across all speakers before input to the training process. 
    The weights were initialized using \textit{Xavier} initialization and \textit{Adam} optimization was used \cite{xavier2010init, diederik2014adam}.
    The training methods of the SD- and SA-WaveNets were similar but the initialization process was different in each case the SD model was initialized by Xavier initialization whereas the SA-WaveNet was initialized using the SI-WaveNet model whose weights were optimized toward the target speaker's database to represent speaker-specific characteristics.

    To construct a baseline TTS acoustic model, we employed a \textit{shared hidden layer (SHL) acoustic model} \cite{fan2015multi, pascual2016multi}.
    The linguistic input feature vectors were 356-dimensional contextual information consisting of 330 binary features of categorical linguistic contexts and 26 features of numerical linguistic contexts. 
    The output vectors consisted of all the acoustic parameters together with their time dynamics \cite{furui1986speaker}.
    Before training, both input and output features were normalized to have zero mean and unit variance. 
    The SHL consisted of three feedforward layers with 1,024 units and one long short-term memory layer with 512 memory blocks. 
    The weights were trained using a \textit{backpropagation through time} algorithm with Adam optimization \cite{williams1990efficient}.

    In the synthesis step, the means of all acoustic features were predicted by the SHL model first, then a speech parameter generation algorithm was applied with the pre-computed global variances \cite{ze2013statistical, tokuda2000speech}.
    To enhance spectral clarity, an LSF-sharpening filter was also applied to the spectral parameters \cite{song2017effective}.
    To reconstruct the speech signal, the generated acoustic features were used to compose the input auxiliary features. By conditioning these features, the WaveNet generated discrete symbols corresponding to the quantized speech signal, and its dynamic was recovered via $\mu$-law expansion.
    
    The setups for training the SI-, SD-, and SA-ExcitNets were the same as those for the WaveNets but the ExcitNet-based framework predicted the distribution of the excitation signal,  obtained by passing the speech signal through the LP analysis filter.
    Similar to the WaveNet vocoder, the ExcitNet vocoder generated the excitation sequence in the synthesis step.
    Ultimately, the speech signal was reconstructed through an LP synthesis filter.

\subsection{Objective test results}
\label{ssec:ch4-2}   
    %%%%%%%%%%%%%%% Table: spss%%%%%%%%%%%%%%%%%%%%%%%%
	\begin{table}[!t]
	\begin{center}
	\caption{LSD (dB) and F0 RMSE (Hz) results for the Korean male (KRM) and the Korean female (KRF) speakers (SPKs): the smallest errors are in bold.}
	\label{table:obj-spss}
	%\vspace*{-8pt}	
	{\small        
	\begin{tabular}{>{\centering}m{.10\linewidth}|c||c|c||c|c}
	\Xhline{2\arrayrulewidth}
    \multicolumn{1}{c|}{\multirow{2}{*}{SPK}}   &\multicolumn{1}{c||}{\multirow{2}{*}{System}}   & \multicolumn{2}{c||}{WaveNet}         & \multicolumn{2}{c}{ExcitNet} \\\cline{3-6}
    \multicolumn{1}{c|}{}  & \multicolumn{1}{c||}{}  & \multicolumn{1}{c|}{LSD} & \multicolumn{1}{c||}{F0 RMSE}  & \multicolumn{1}{c|}{LSD} & \multicolumn{1}{c}{F0 RMSE}  \\ \hline\hline
	    & SD  & 4.37         & 21.30          & 3.93          & 14.83        \\	
	KRM & SI  & 4.06         & 14.76          & 3.86          & 14.39         \\	
	    & SA  &\textbf{4.03} &\textbf{14.16}	&\textbf{3.82}	&\textbf{14.03} \\ \hline\hline
	    & SD  & 4.78         & 48.75          & 4.50          & 39.14        \\	
	KRF & SI  & 4.51         & 35.53          & 4.42          & 36.28         \\	
	    & SA  &\textbf{4.45} &\textbf{35.45}	&\textbf{4.36}	&\textbf{35.47} \\
			\Xhline{2\arrayrulewidth}
	\end{tabular}}
	\end{center}
	%\vspace*{-12pt} 
	\end{table}
    %%%%%%%%%%%%%%% Table: spss %%%%%%%%%%%%%%%%%%%%%%%%
    
    %%%%%%%%%%%%%%% Table: spss%%%%%%%%%%%%%%%%%%%%%%%%
	\begin{table}[!t]
	\begin{center}
	\caption{Objective test results in the large-scale (7 hours) adaptation: the smallest errors are in bold.}
	\label{table:obj-spss-lsa}
	%\vspace*{-8pt}	
	{\small        
	\begin{tabular}{>{\centering}m{.10\linewidth}|c||c|c||c|c}
	\Xhline{2\arrayrulewidth}
    \multicolumn{1}{c|}{\multirow{2}{*}{SPK}}   &\multicolumn{1}{c||}{\multirow{2}{*}{System}}   & \multicolumn{2}{c||}{WaveNet}         & \multicolumn{2}{c}{ExcitNet} \\\cline{3-6}
    \multicolumn{1}{c|}{}  & \multicolumn{1}{c||}{}  & \multicolumn{1}{c|}{LSD} & \multicolumn{1}{c||}{F0 RMSE}  & \multicolumn{1}{c|}{LSD} & \multicolumn{1}{c}{F0 RMSE}  \\ \hline\hline
	    & SD  & 3.63         & 11.28          & 3.37          & 12.08        \\	
	KRM & SI  & 3.66         & 12.08          & 3.40          & 11.65         \\	
	    & SA  &\textbf{3.58} &\textbf{11.04}	&\textbf{3.33}	&\textbf{10.93} \\ \hline\hline
	    & SD  & 4.08         & 28.75          & 3.95          & 28.76        \\	
	KRF & SI  & 4.13         & 28.77          & 4.00          & 28.84         \\	
	    & SA  &\textbf{4.03} &\textbf{28.57}	&\textbf{3.90}	&\textbf{28.36} \\
			\Xhline{2\arrayrulewidth}
	\end{tabular}}
	\end{center}
	\vspace*{-12pt} 
	\end{table}
    %%%%%%%%%%%%%%% Table: spss %%%%%%%%%%%%%%%%%%%%%%%%

\subsubsection{Parametric text-to-speech}
\label{ssec:ch4-2-1}   
    To verify the performance of the proposed method, we measured distortions between the original speech and the synthesized speech with log-spectral distance (LSD; dB) and F0 root mean square error (RMSE; Hz) measures. 
    Table~\ref{table:obj-spss} presents the test results with respect to the different types of training methods.
    The findings can be outlined as follows: (1) The proposed SA training method reconstructs more accurate speech signals than the SD and SI models in both WaveNet and ExcitNet vocoders.
    (2) Among the different vocoding systems, the ExcitNet-based framework performed better than the WaveNet-based one in terms of spectral distortion because the adoption of an adaptive spectral filter for the ExcitNet vocoder is beneficial for more accurately modeling the target speech signals. 
    
    To verify the effectiveness of the proposed algorithm in a \textit{large amount} of training database condition, additional experiments were conducted by changing the adaptation data size from 10 minutes to 7 hours.
    For the comparison, the amount of database to train the SD model and the SHL acoustic model was also increased to 7 hours.
    Table~\ref{table:obj-spss-lsa} shows the test results, which confirms that adapting the vocoding model toward target speaker's database is still advantageous to improve the modeling accuracy regardless of the amount of training data set.

%%%%%%%%%%%%%%%%%%% Fig: ExcitNet %%%%%%%%%%%%%%%%%%%%%%%%%%%%%%
	\begin{figure}[!t]
	\begin{minipage}[t]{.99\linewidth}
	\centerline{\epsfig{figure=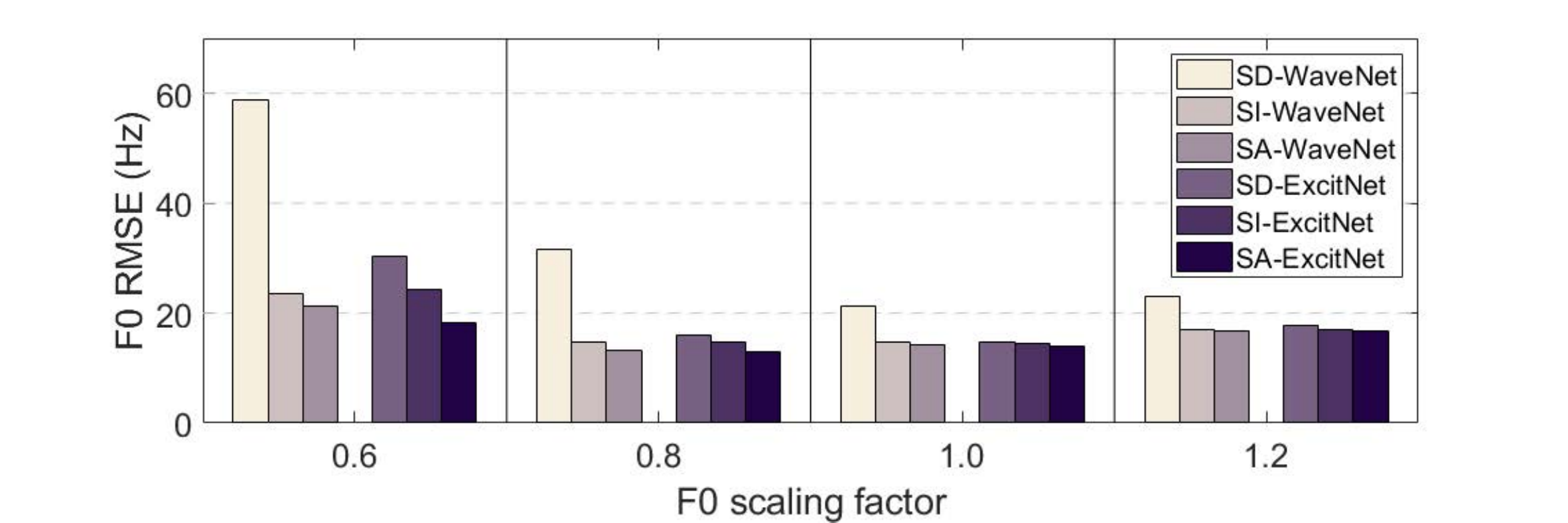,width=95mm}}
	%\vspace*{-2pt}	
	\centerline{(a)}  \medskip
	%\vspace*{-6pt}	
	\end{minipage}
	\begin{minipage}[t]{.99\linewidth}
	\centerline{\epsfig{figure=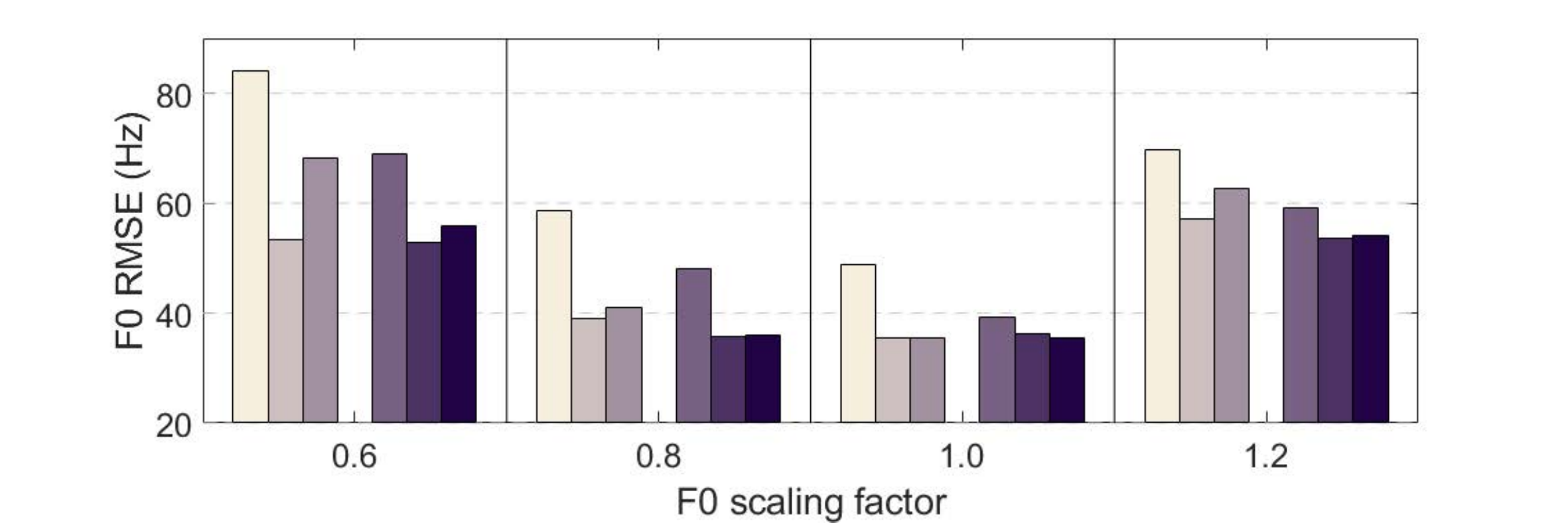,width=95mm}}
	%\vspace*{-2pt}	
	\centerline{(b)}  \medskip
	\end{minipage}	
	%\vspace*{-6pt} 
	\caption{F0 RMSE (Hz) results with respect to different values of the scaling factor: (a) Korean male and (b) Korean female speakers.}	
%	\vspace*{-8pt} 
	\label{fig:f0_mod}
	\end{figure}
%%%%%%%%%%%%%%%%%%% Fig: ExcitNet %%%%%%%%%%%%%%%%%%%%%%%%%%%%%%
    
\subsubsection{Speech modification}
\label{ssec:ch4-2-2}   

    To further verify the effectiveness of the proposed SA training method, we investigated the performance variation of neural vocoders when F0 is manually modified. It has already been shown that the SI model effectively generates pitch-modified synthesized speech \cite{hayashi2017investigation}.
    As the entire network of the SA approach in the present study was adapted from an SI model, it was expected to further improve performance compared to conventional SD approaches.
    
    In this experiment, the F0 trajectory was first generated by the TTS acoustic model and then multiplied by a scaling factor ($0.6, 0.8, 1.0,$ and $1.2$) to modify the auxiliary feature vectors. 
    Finally, the speech signal was synthesized using the neural vocoding systems. 
    Figure~\ref{fig:f0_mod} illustrates the F0 RMSE (Hz) test results with respect to the different values of scaling factor.
    The results can be analyzed as follows: (1) The proposed SA training models result in smaller modification errors than the conventional SD approaches.
    (2) The performance of the SI and SA methods was not much different, but the SI method was somewhat better than the SA method for the female speaker case, especially when the modification ratio was high. 
    (3) In all experiments, the ExcitNet-based system performed better than the WaveNet-based one because the ExcitNet model was instructed to learn the variation of vocal cord movement.

%%%%%%%%%%%%%%%%%%% Fig: ExcitNet %%%%%%%%%%%%%%%%%%%%%%%%%%%%%%
	\begin{figure}[!t]
	\begin{minipage}[t]{.99\linewidth}
	\centerline{\epsfig{figure=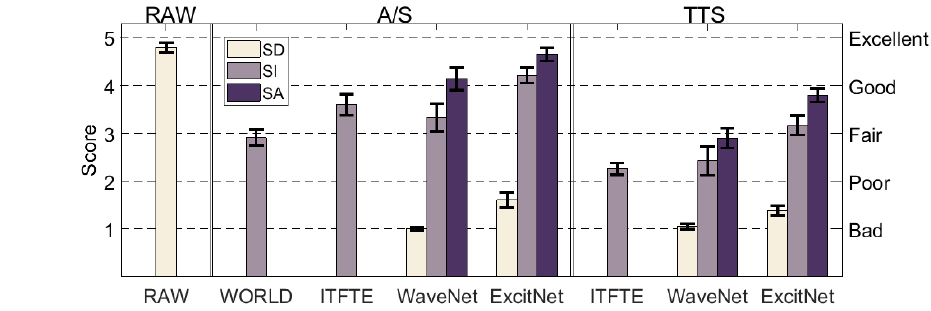,width=95mm}}
	%\vspace*{-2pt}	
	\centerline{(a)}  \medskip
	%\vspace*{-6pt}	
	\end{minipage}
	\begin{minipage}[t]{.99\linewidth}
	\centerline{\epsfig{figure=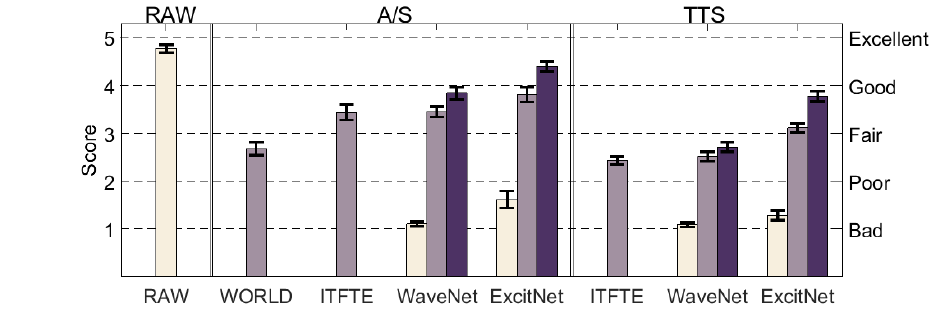,width=95mm}}
	%\vspace*{-2pt}	
	\centerline{(b)}  \medskip
	\end{minipage}	
	%\vspace*{-6pt} 
	\caption{MOS results with 95\% confidence intervals. Acoustic features extracted from recorded speech and generated from an acoustic model were used to compose the input auxiliary features in the A/S and TTS tasks, respectively: (a) Korean male and (b) Korean female speakers.}	
	\vspace*{-8pt} 
	\label{fig:mos}
	\end{figure}
%%%%%%%%%%%%%%%%%%% Fig: ExcitNet %%%%%%%%%%%%%%%%%%%%%%%%%%%%%%
 
\subsection{Subjective test results}
\label{ssec:ch4-3} 

    To evaluate the perceptual quality of the proposed system, mean opinion score (MOS) tests were performed\footnote{Generated audio samples are available at the following url:\\ \url{https://sewplay.github.io/demos/vocoder_adaptation}}.
    In the tests, twelve native Korean listeners were asked to make quality judgments about the synthesized speech based on the following five possible responses: 1 = Bad; 2 = Poor; 3 = Fair; 4 = Good; and 5 = Excellent. 
    Note that the listening tests were performed in an acoustically isolated room using a Sennheiser HD650 headphone.
    In total, twenty utterances were randomly selected from the test set and were then synthesized by using the different neural vocoders. 
    To verify vocoding performance, the speech samples synthesized by the conventional vocoders such as ITFTE and WORLD (D4C edition \cite{morise2016d4c}) were also included.

    As presented in Figure~\ref{fig:mos}, the subjective test results confirm the effectiveness of each system in several ways: 
    (1) In both the analysis and synthesis (A/S) and TTS frameworks, the SD vocoders performed worst because it was difficult to learn the target speaker's characteristics with such a small amount of training data. 
    (2) As the SI models could represent multiple speaker's voices, they were able to synthesize more natural speech than than the SD approaches.  
    (3) Across all the training methods, the SA version achieved the best quality, which confirms that adapting the multi-speaker model to the target speaker's database is beneficial for the vocoding performance. 
    (4) Comparing with the WaveNet, the ExcitNet performed better overall, confirming that decoupling the formant component of the speech signal via an LP inverse filter significantly improves the modeling accuracy. 
    (5) Consequently, the TTS system with the proposed SA-ExcitNet vocoder achieved 3.80 and 3.77 MOS for the Korean male and Korean female speakers, respectively.

\section{Conclusion}%\vspace{-2mm}
    This paper proposed speaker-adaptive neural vocoders for parametric TTS systems when the amount of target speaker's data is insufficient. 
    Using an initial speaker-independent trained model, the system first captured universal attributes from the waveform of multiple speakers'.
    This model was then fine-tuned with the target speaker's database to successfully represent speaker-specific characteristics using only ten minutes of training data. Adapting an ExcitNet framework with spectral filters also helped to improve the modeling accuracy.
    The experimental results verified that the TTS system with the proposed speaker-adaptive neural vocoder performed significantly better than traditional versions with linear predictive coding-based vocoders and systems with similarly configured neural vocoders trained both speaker-dependently and speaker-independently. 
    Future research includes integrating the entire framework into speech synthesis systems that use an end-to-end approach.

\bibliographystyle{IEEEtran}
\bibliography{mybib}

\begin{thebibliography}{10}
\providecommand{\url}[1]{#1}
\csname url@samestyle\endcsname
\providecommand{\newblock}{\relax}
\providecommand{\bibinfo}[2]{#2}
\providecommand{\BIBentrySTDinterwordspacing}{\spaceskip=0pt\relax}
\providecommand{\BIBentryALTinterwordstretchfactor}{4}
\providecommand{\BIBentryALTinterwordspacing}{\spaceskip=\fontdimen2\font plus
\BIBentryALTinterwordstretchfactor\fontdimen3\font minus
  \fontdimen4\font\relax}
\providecommand{\BIBforeignlanguage}[2]{{%
\expandafter\ifx\csname l@#1\endcsname\relax
\typeout{** WARNING: IEEEtran.bst: No hyphenation pattern has been}%
\typeout{** loaded for the language `#1'. Using the pattern for}%
\typeout{** the default language instead.}%
\else
\language=\csname l@#1\endcsname
\fi
#2}}
\providecommand{\BIBdecl}{\relax}
\BIBdecl

\bibitem{van2016conditional}
A.~Van Den~Oord, N.~Kalchbrenner, L.~Espeholt, O.~Vinyals, A.~Graves
  \emph{et~al.}, ``Conditional image generation with {P}ixel{CNN} decoders,''
  in \emph{Proc. NIPS}, 2016, pp. 4790--4798.

\bibitem{van2016wavenet}
A.~Van Den~Oord, S.~Dieleman, H.~Zen, K.~Simonyan, O.~Vinyals, A.~Graves,
  N.~Kalchbrenner, A.~Senior, and K.~Kavukcuoglu, ``Wave{N}et: {A} generative
  model for raw audio,'' \emph{CoRR abs/1609.03499}, 2016.

\bibitem{tamamori2017speaker}
A.~Tamamori, T.~Hayashi, K.~Kobayashi, K.~Takeda, and T.~Toda,
  ``Speaker-dependent {W}ave{N}et vocoder,'' in \emph{Proc. INTERSPEECH}, 2017,
  pp. 1118--1122.

\bibitem{hayashi2017investigation}
T.~Hayashi, A.~Tamamori, K.~Kobayashi, K.~Takeda, and T.~Toda, ``An
  investigation of multi-speaker training for wavenet vocoder,'' in \emph{Proc.
  ASRU}, 2017, pp. 712--718.

\bibitem{hu2017ustc}
Y.-J. Hu, C.~Ding, L.-J. Liu, Z.-H. Ling, and L.-R. Dai, ``The {USTC} system
  for blizzard challenge 2017,'' in \emph{Proc. Blizzard Challenge Workshop},
  2017.

\bibitem{shen2017natural}
J.~Shen, R.~Pang, R.~J. Weiss, M.~Schuster, N.~Jaitly, Z.~Yang, Z.~Chen,
  Y.~Zhang, Y.~Wang, R.~Skerry-Ryan \emph{et~al.}, ``Natural {TTS} synthesis by
  conditioning {W}ave{N}et on mel spectrogram predictions,'' in \emph{Proc.
  ICASSP}, 2018, pp. 4779--4783.

\bibitem{adiga2018use}
N.~Adiga, V.~Tsiaras, and Y.~Stylianou, ``On the use of {W}ave{N}et as a
  statistical vocoder,'' in \emph{Proc. ICASSP}, 2018, pp. 5674--5678.

\bibitem{song2019excitnet}
E.~Song, K.~Byun, and H.-G. Kang, ``Excit{N}et vocoder: {A} neural excitation
  model for parametric speech synthesis systems,'' in \emph{Proc. EUSIPCO},
  2019, pp. 1179--1183.

\bibitem{yoshimura2018mel}
T.~Yoshimura, K.~Hashimoto, K.~Oura, Y.~Nankaku, and K.~Tokuda,
  ``Mel-cepstrum-based quantization noise shaping applied to
  neural-network-based speech waveform synthesis,'' \emph{IEEE/ACM Trans.
  Audio, Speech, and Lang. Process.}, vol.~26, no.~7, pp. 1173--1180, 2018.

\bibitem{juvela2018speaker}
L.~Juvela, V.~Tsiaras, B.~Bollepalli, M.~Airaksinen, J.~Yamagishi, and P.~Alku,
  ``Speaker-independent raw waveform model for glottal excitation,'' in
  \emph{Proc. INTERSPEECH}, 2018, pp. 2012--2016.

\bibitem{hwang2018lp}
M.-J. Hwang, F.~Soong, E.~Song, X.~Wang, H.~Kang, and H.-G. Kang,
  ``{LP}-{W}ave{N}et: {L}inear prediction-based {W}ave{N}et speech synthesis,''
  \emph{arXiv preprint arXiv:1811.11913}, 2018.

\bibitem{valin2018lpcnet}
J.-M. Valin and J.~Skoglund, ``{LPC}net: {I}mproving neural speech synthesis
  through linear prediction,'' in \emph{Proc. ICASSP}, 2019, pp. 5891--5895.

\bibitem{tachibana2018investigation}
K.~Tachibana, T.~Toda, Y.~Shiga, and H.~Kawai, ``An investigation of noise
  shaping with perceptual weighting for {W}ave{N}et-based speech generation,''
  in \emph{Proc. ICASSP}, 2018, pp. 5664--5668.

\bibitem{wang2018comparison}
X.~Wang, J.~Lorenzo-Trueba, S.~Takaki, L.~Juvela, and J.~Yamagishi, ``A
  comparison of recent waveform generation and acoustic modeling methods for
  neural-network-based speech synthesis,'' in \emph{Proc. ICASSP}, 2018, pp.
  4804--4808.

\bibitem{song2017effective}
E.~Song, F.~K. Soong, and H.-G. Kang, ``Effective spectral and excitation
  modeling techniques for {LSTM}-{RNN}-based speech synthesis systems,''
  \emph{IEEE/ACM Trans. Audio, Speech, and Lang. Process.}, vol.~25, no.~11,
  pp. 2152--2161, 2017.

\bibitem{mehri2017samplernn}
S.~Mehri, K.~Kumar, I.~Gulrajani, R.~Kumar, S.~Jain, J.~Sotelo, A.~Courville,
  and Y.~Bengio, ``{SampleRNN}: {An} unconditional end-to-end neural audio
  generation model,'' in \emph{Proc. ICLR}, 2017.

\bibitem{kalchbrenner2018efficient}
N.~Kalchbrenner, E.~Elsen, K.~Simonyan, S.~Noury, N.~Casagrande, E.~Lockhart,
  F.~Stimberg, A.~van~den Oord, S.~Dieleman, and K.~Kavukcuoglu, ``Efficient
  neural audio synthesis,'' in \emph{Proc. ICML}, 2018, pp. 2410--2419.

\bibitem{prenger2019waveglow}
R.~Prenger, R.~Valle, and B.~Catanzaro, ``Wave{G}low: {A} flow-based generative
  network for speech synthesis,'' in \emph{Proc. ICASSP}, 2019, pp. 3617--3621.

\bibitem{xavier2010init}
X.~Glorot and Y.~Bengio, ``Understanding the difficulty of training deep
  feedforward neural networks,'' in \emph{Proc. AISTATS}, 2010, pp. 249--256.

\bibitem{diederik2014adam}
\BIBentryALTinterwordspacing
D.~P. Kingma and J.~Ba, ``Adam: {A} method for stochastic optimization,''
  \emph{CoRR}, vol. abs/1412.6980, 2014. [Online]. Available:
  \url{http://arxiv.org/abs/1412.6980}
\BIBentrySTDinterwordspacing

\bibitem{fan2015multi}
Y.~Fan, Y.~Qian, F.~K. Soong, and L.~He, ``Multi-speaker modeling and speaker
  adaptation for {DNN}-based {TTS} synthesis,'' in \emph{Proc. ICASSP}, 2015,
  pp. 4475--4479.

\bibitem{pascual2016multi}
S.~Pascual and A.~Bonafonte, ``Multi-output {RNN}-{LSTM} for multiple speaker
  speech synthesis and adaptation,'' in \emph{Proc. EUSIPCO}, 2016, pp.
  2325--2329.

\bibitem{furui1986speaker}
S.~Furui, ``Speaker-independent isolated word recognition using dynamic
  features of speech spectrum,'' \emph{IEEE Trans. Acoust., Speech Signal
  Process.}, vol.~34, no.~1, pp. 52--59, 1986.

\bibitem{williams1990efficient}
R.~J. Williams and J.~Peng, ``An efficient gradient-based algorithm for on-line
  training of recurrent network trajectories,'' \emph{Neural computat.},
  vol.~2, no.~4, pp. 490--501, 1990.

\bibitem{ze2013statistical}
H.~Zen, A.~Senior, and M.~Schuster, ``Statistical parametric speech synthesis
  using deep neural networks,'' in \emph{Proc. ICASSP}, 2013, pp. 7962--7966.

\bibitem{tokuda2000speech}
K.~Tokuda, T.~Yoshimura, T.~Masuko, T.~Kobayashi, and T.~Kitamura, ``Speech
  parameter generation algorithms for {HMM}-based speech synthesis,'' in
  \emph{Proc. ICASSP}, 2000, pp. 1315--1318.

\bibitem{morise2016d4c}
M.~Morise, ``{D4C}, a band-aperiodicity estimator for high-quality speech
  synthesis,'' \emph{Speech commun.}, vol.~84, pp. 57--65, 2016.

\end{thebibliography}

\end{document}